\documentclass[sigconf]{acmart}

\author{Jianfeng Deng}
\affiliation{%
  \institution{Guangxi University}
  \city{Nanning}
  \country{China}
}
\email{jianfeng_web@163.com}

\author{Qingfeng Chen}
\authornote{Corresponding author}
\affiliation{%
  \institution{Guangxi University}
  \city{Nanning}
  \country{China}
}
\email{qingfeng@gxu.edu.cn}

\author{Debo Cheng}
\authornotemark[1]
\affiliation{%
  \institution{University of South Australia}
  \city{Adelaide}
  \country{Australia}
}
\email{debo.cheng@unisa.edu.au}

\author{Jiuyong Li}
\affiliation{%
  \institution{University of South Australia}
  \city{Adelaide}
  \country{Australia}
}
\email{jiuyong.li@unisa.edu.au}

\author{Lin Liu}
\affiliation{%
  \institution{University of South Australia}
  \city{Adelaide}
  \country{Australia}
}
\email{lin.liu@unisa.edu.au}

\author{Shichao Zhang}
\affiliation{%
  \institution{Guangxi Normal University}
  \city{Nanning}
  \country{China}
}
\email{zhangsc@mailbox.gxnu.edu.cn}

\renewcommand\footnotetextcopyrightpermission[1]{} 

\DeclareMathAlphabet{\mathcal}{OMS}{cmsy}{m}{n}  

\usepackage{hyperref}
\usepackage{url}
\usepackage{graphicx}
\usepackage{caption}

\usepackage{diagbox}
\usepackage{multirow}
\usepackage{booktabs}
\usepackage{array}
\usepackage{enumitem}

\usepackage{amsthm}

\usepackage{amssymb} 
\usepackage{amsmath}

\usepackage{amsfonts}

\newtheorem{assumption}{Assumption}		%
\newtheorem{theorem}{Theorem}		%
\newtheorem{definition}{Definition}

\usepackage[ruled,linesnumbered]{algorithm2e}

\usepackage{cleveref}

\crefname{equation}{equation}{equations}
\crefformat{equation}{(#2#1#3)}

\begin{document}

\title{A Novel Generative Model with Causality Constraint for Mitigating Biases in Recommender Systems}

\begin{abstract}
Accurately predicting counterfactual user feedback is essential for building effective recommender systems. However, latent confounding bias can obscure the true causal relationship between user feedback and item exposure, ultimately degrading recommendation performance. Existing causal debiasing approaches often rely on strong assumptions—such as the availability of instrumental variables (IVs) or strong correlations between latent confounders and proxy variables—that are rarely satisfied in real-world scenarios.
To overcome these limitations, we propose a novel generative framework called \underline{L}atent \underline{C}ausality \underline{C}onstraints for \underline{D}ebiasing representation learning in \underline{R}ecommender Systems (LCDR). Specifically, LCDR leverages an identifiable Variational Autoencoder (iVAE) as a causal constraint to align the latent representations learned by a standard Variational Autoencoder (VAE) through a unified loss function. This alignment allows the model to leverage even weak or noisy proxy variables to recover latent confounders effectively. The resulting representations are then used to improve recommendation performance.
Extensive experiments on three real-world datasets demonstrate that LCDR consistently outperforms existing methods in both mitigating bias and improving recommendation accuracy.

\end{abstract}


\maketitle

\section{Introduction}
Recommender systems are widely used in various real-world domains, such as video streaming~\cite{5}, e-commerce~\cite{36}, and web search~\cite{32,li2024ripple}. Numerous recommendation algorithms have been developed, including Matrix Factorisation (MF)~\cite{12}, Factorisation Machine (FM)~\cite{fm}, Deep Factorisation Machine (DeepFM)~\cite{deepfm}, Deep Cross Network (DCN)~\cite{dcn}, Deep Interest Network (DIN)~\cite{din}, and LightGCN~\cite{lightgcn}.
Nevertheless, these approaches often depend on statistical correlations, which can introduce various estimation biases, such as popularity bias~\cite{popularity1}, selection bias~\cite{select1,15}, and conformity bias~\cite{conformity1}.
To tackle these challenges, causality-driven recommender systems have gained attention as a promising solution, seeking to improve fairness and precision by reducing these biases~\cite{21,zhang2025mitigating}.

Causal inference-based recommendation methods have shown effectiveness in addressing various biases introduced by confounders in recommender systems. 
Recently, a large number of causality-based recommendation algorithms have been developed to address these biases. 
Some methods rely on traditional causal inference techniques, such as inverse probability weighting (IPW)~\cite{23} and backdoor adjustment~\cite{20}. For instance, Inverse Propensity Scoring (IPS)~\cite{23} uses propensity scores and IPW to mitigate selection bias. D2Q~\cite{34} employs backdoor adjustment to reduce bias caused by video duration. Popularity De-biasing Algorithm (PDA)~\cite{35} reweights training samples through a causal model to eliminate the interference of popularity in the target variable. 
AKBDR-Gau~\cite{li2024debiased} adopts an adaptive kernel balancing methodology to learn an optimal balancing function that effectively mitigates selection bias.
Multifactorial IPS (Mulfact-IPS)~\cite{huang2024going} introduces a propensity estimation approach to mitigate multi-factorial bias, which arises from both item-specific and rating value factors.
However, these methods typically rely on specific causal relationship assumptions and often overlook latent confounders that may exist in real-world scenarios.

Recent advances in causal recommendation approaches emphasize the importance of addressing latent confounders~\cite{idcf,23}. Although these latent confounders are inherently unmeasurable, they can significantly distort the output of recommendation systems~\cite{20,cheng2022local,cheng2024data}. For instance, high-spending user groups often exhibit distinct consumption patterns: they tend to prefer premium-priced products but may assign lower ratings due to their stringent evaluation criteria. As a result, recommendation systems might misinterpret this group’s preferences for high-end products, thereby impacting recommendation accuracy.

Several technical solutions have been developed to mitigate such biases.
For example, Robust Deconfounder IPS (RD-IPS)~\cite{4} assumes that the impact of latent confounders on exposure is limited and corrects for confounders in the estimation of the propensity score through an adversarial training mechanism. 
Invariant Preference Learning (InvPref)~\cite{30} addresses confounding bias based on the assumption of environmental invariance, while IV4Rec~\cite{IV-serach-data} leverages instrumental variable methods, which require satisfying the independence assumption of instrumental variables, to eliminate confounding effects. ~\cite{li2023balancing} relies on unbiased data to address  confounding bias, but truly unbiased data are difficult to collect in practice. These works~\cite{29,38} assume that the inference of latent confounders can be accomplished without the need to introduce additional proxy variables (proxy variables refer to observable substitute indicators for latent confounders that cannot be directly measured).
iDCF~\cite{idcf} achieves an identifiable modeling of latent variables under ideal assumptions, such as the latent variable and the proxy variable being highly correlated. 
However, these ideal assumptions are often difficult to fully satisfy in real-world applications.

To address the aforementioned issues, we propose a novel generative model, named \underline{L}atent \underline{C}ausality \underline{C}onstraints for \underline{D}ebiasing \underline{R}ecommender Systems (LCDR). Specifically, LCDR leverages an identifiable Variational Autoencoder (iVAE) as a causal constraint to align the learned latent representations of a standard Variational Autoencoder (VAE) through a unified loss function. 
By indirectly influencing the latent representations through proxy variables, this loss function enables robust recovery of latent structures, even when the proxies are noisy or of low quality in practice. The learned representations are subsequently utilised to enhance the performance of the recommendation model. We conducted extensive experiments on three real-world datasets, and the results demonstrate that LCDR significantly outperforms existing methods in mitigating bias and improving recommendation accuracy.
In summary, the main contributions of our work include:
\begin{itemize}
    \item We propose a novel generative model, Latent Causality Constraints for Debiasing representation learning Recommender Systems (LCDR), which enables the generated representations to more accurately capture the distributional characteristics of real-world data, thereby improving adaptability to practical scenarios.
    \item We introduce an innovative causality constraint that indirectly adjusts the latent variables via the proxy variables, significantly enhancing the model’s robustness and applicability in practical applications.
    \item Extensive experiments conducted on three public real-world datasets demonstrate that LCDR outperforms state-of-the-art recommendation methods in mitigating bias and improving recommendation accuracy.
\end{itemize}

\section{Related Work}
This section introduces the recent works on recommender systems, categorised into two primary groups: correlation-oriented and causality-oriented systems.

\paragraph{Traditional Recommendation Algorithms.}
Traditional recommendation methods include collaborative filtering-based approaches (e.g., Matrix Factorisation (MF) \cite{12}), neural network-based recommendation algorithms (e.g., Deep Factorisation Machine (DeepFM) \cite{deepfm}, Deep Cross Network (DCN) \cite{dcn}, Deep Interest Network (DIN) \cite{din}), and graph neural network-based algorithms (e.g., LightGCN \cite{lightgcn}). While these conventional methods have demonstrated effectiveness, they often introduce biases in recommendation results. In this study, we primarily focus on causal inference frameworks for recommendation systems.

\begin{figure}
    \centering{}
    \includegraphics[width=0.45\textwidth]{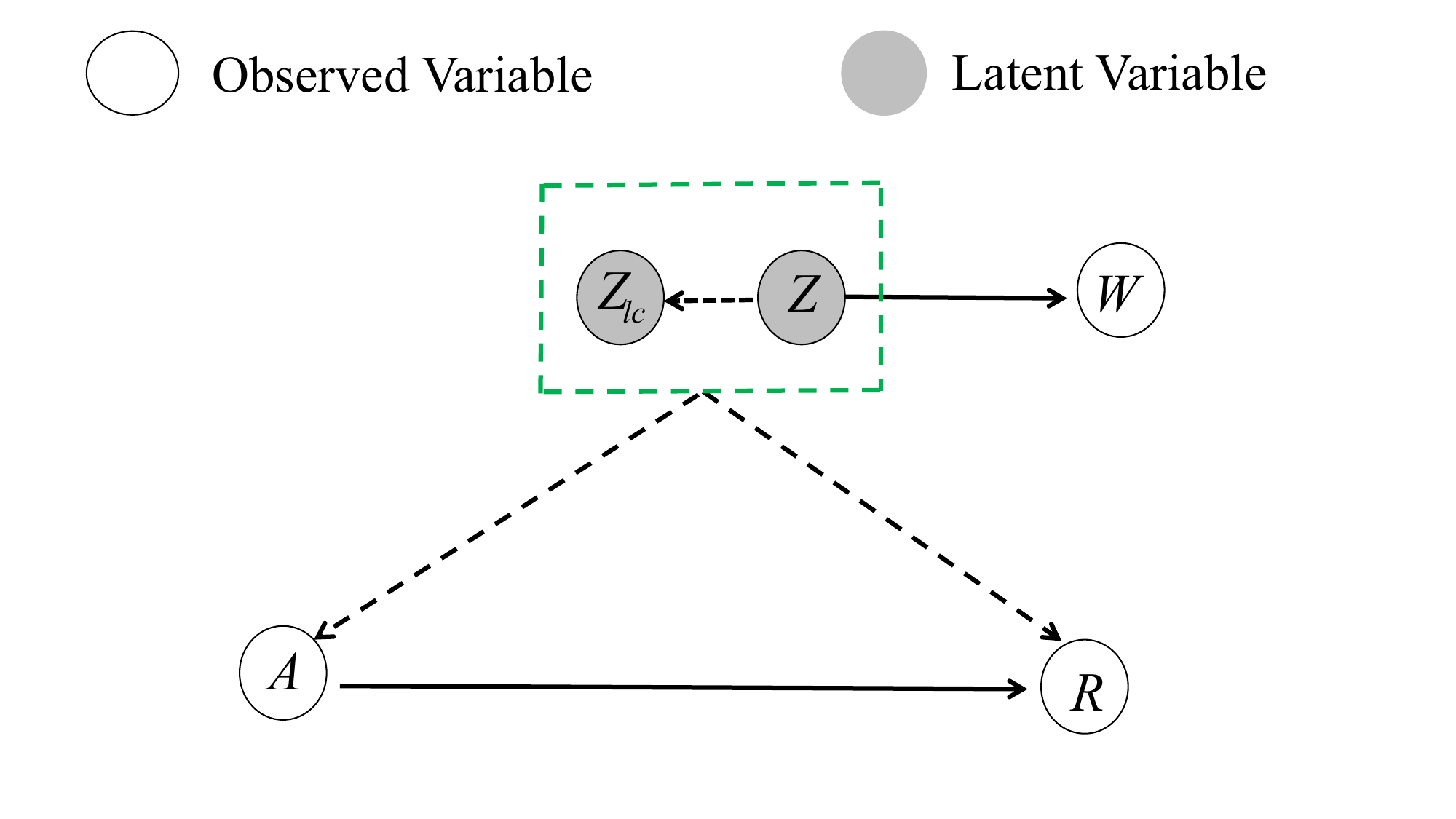} \vspace{-20pt}
    \caption{An illustrative DAG showing how latent confounders affect recommendation systems. ${R}$ represents outcome variables,  ${A}$ indicates exposure status, ${Z}$ denotes latent causal representations learned by iVAE,  and ${Z_{lc}}$ represents Latent representations learned by latent causal-constrained VAE. ${W}$ represents the proxy variables, ${W}$ does not directly adjust the distribution of ${Z_{lc}}$ but indirectly adjusts the distribution of ${Z_{lc}}$ through ${Z}$.}
    \label{fig:2}
\vspace{-5pt}
\end{figure}

\paragraph{Causal Recommendation Algorithms for Addressing Specific Biases.}
Numerous debiasing approaches in recommendation systems leverage traditional causal inference techniques, such as IPW and backdoor adjustment. For instance, IPS~\cite{23} employs propensity scores and IPW to mitigate selection bias, D2Q~\cite{34} addresses video duration bias using backdoor adjustment, and PDA~\cite{35} alleviates popularity bias by incorporating causal interventions during model training. PPAC~\cite{ning2024debiasing} employs counterfactual inference techniques to mitigate Global Popularity (GP) bias. 
DR-JL~\cite{wang2019doubly} tackles the issue of missing not at random (MNAR) data in recommender systems by applying a doubly robust estimation approach.
AKBDR-Gau~\cite{li2024debiased} utilises an adaptive kernel balancing strategy to determine the optimal balancing function that reduces selection bias.
Mulfact-IPS~\cite{huang2024going} introduces a multifactorial approach to tackle biases arising from both item-specific and rating-value factors.
Although these methods have shown promising results, they often overlook latent confounders, which can significantly impact the accuracy of traditional causal estimations~\cite{cheng2023causal}. 

Recent studies have proposed methods to address confounding bias caused by latent confounders, often relying on idealised assumptions~\cite{4,30,33,37,chen2019ildmsf,lan2023multiview,29,idcf}.  For example, RD-IPS~\cite{4} assumes that the latent confounders have a limited effect on exposure and mitigates their impact through sensitivity analysis and adversarial learning. InvPref~\cite{30} addresses confounding bias based on the assumption of environment invariance, while IV4Rec~\cite{IV-serach-data} removes confounding effects using instrumental variable methods.
Methods such as ~\cite{li2023balancing} rely on unbiased data to address  confounding bias, but truly unbiased data are difficult to collect in practice.
Methods such as ~\cite{29,38} assume that confounders can be inferred without introducing additional proxy variables. Similarly, iDCF~\cite{idcf} achieves identifiable modeling of latent variables under strong assumptions, such as a high correlation between the latent variable and its proxy. However, these assumptions are often overly idealistic and difficult to satisfy in real-world scenarios.

Compared to the reviewed works above, our proposed LCDR method introduces the causal constraints that are more closely aligned with real-world conditions, significantly enhancing the robustness and practical applicability of recommendation models.

\begin{table}[t]
  \centering
  \caption{Definitions and Notations}
  \label{tab:notation}
  \begin{tabular}{c p{5cm}}
    \toprule
    \textbf{Symbol} & \textbf{Definition} \\
    \midrule
    $\mathcal{U}$ & The set of users \\
    $\mathcal{I}$ & The set of Items \\
    ${A}$ & Exposure vector. \\
    ${R}$ & Observational feedback from user $u$. \\
    ${Z}$ & Identifiable latent causal representations learned by iVAE.  \\
    ${Z_{vae}}$ & Latent representations learned by VAE. \\
    ${Z_{lc}}$ & Latent representations learned by latent causal-constrained VAE. \\
    ${W}$ & The set of proxy variables. \\
    $r_{ui}$ & Observed outcome (feedback from user-item pairs). \\
    $r_{ui}^a$ & The potential outcome of user $u$ when item $i$ is exposure as $A=a$. \\
    \bottomrule
  \end{tabular}
\end{table}

\section{Notations and Problem Definition}

\subsection{Notations}

We have outlined the basic notations used in this work, as presented in Table~\ref{tab:notation}. The main symbols are defined as follows: vectors are represented by capital letters without boldface (e.g., ${A}$), and their individual components are denoted using lowercase letters with subscripts (e.g., $a_{ui}$).

Let $\mathcal{U}$ represent the set of users, with $|\mathcal{U}|=m$ denoting the total number of users. Similarly, let $\mathcal{I}$ represent the set of items, with $|\mathcal{I}|=n$ indicating the total number of items. For each user $u \in \mathcal{U}$, the exposure vector ${A}$ is defined as ${A} = [a_{u1}, a_{u2}, \dots, a_{un}] \in \{0,1\}^n$, where $a_{ui} = 1$ means that item $i$ is shown to user $u$, and $a_{ui} = 0$ means it is not. The feedback vector from user $u$ is denoted by ${R} = [r_{u1}, r_{u2}, \dots, r_{un}]$, where each $r_{ui}$ represents the feedback received from user $u$ for item $i$.

 In this work, we utilise the potential outcomes framework to model our method.  Let $r_{ui}^a$ denote the potential outcome associated with exposure status ${A}=a$. Previous research \cite{29} assumes that the exposure of item $i$ to user $u$ is the sole factor influencing $r_{ui}^a$. Furthermore, we employ the set of proxy variables ${W}$ to learn the latent causal representations ${Z}$ from the iVAE model. Simultaneously, we leverage ${Z}$ to constrain the representations generated by the VAE model. We define the representations constrained by the latent causal representations as ${Z_{lc}}$.

\subsection{Problem Definition}

A recommendation system aims to estimate the probability that a user $u \in \mathcal{U}$  will give positive feedback on an item $i \in \mathcal{I}$ when the item is exposed, represented as $p(r_{ui} = 1 \mid a_{ui} = 1)$.
In real-world scenarios, recommendation systems often encounter the issue of \textbf{latent confounders},  $Z_{lc}$, which are unobserved variables that influence both the feedback $r_{ui}$ and the exposure status $A$. These latent confounders introduce bias, ultimately compromising the accuracy of the recommendations.

In Figure \ref{fig:2}, $W$ denotes the proxy variable, which is widely present in publicly available recommendation system datasets. For instance, MovieLens includes user demographics and movie genres; The Coat dataset including gender, age; The Amazon Dataset provides product categories and purchase sequences; Yelp contains business locations and review timestamps.
$Z$ denotes the latent representation recovered by iVAE under the ideal assumption using noise-free proxy variables, while $Z_{lc}$ denotes the latent representation indirectly recovered under real-world scenarios with noisy proxy variables.

Since the true confounding structure is unobserved, accurately identifying it without proxies is challenging~\cite{9}. For instance, representations like $Z_{vae}$, learned by VAE without proxy variables, often fail to capture the true latent confounding structure. In contrast, models such as iVAE use proxy variables under the ideal assumption of strong correlation with latent confounders to enable identifiability.
However, in practice, proxy variables are often of low quality and fail to satisfy such rigorous assumptions, limiting the applicability of existing methods.
Thus, the formal problem definition of our work is stated as follows:

\begin{definition}
The objective is to estimate the conditional distribution \( P(r_{ui} \mid a_{ui}, Z_{lc}) \) by inferring the latent confounder \( Z_{lc} \) from the low-quality proxy variable \( W \) and item exposure \( A \). To achieve this, we propose an indirect adjustment approach that leverages low-quality proxy variables: by regulating the structure of identifiable latent confounders \( Z \), the proxy variable \( W \) indirectly guides the structural optimisation of latent confounders $Z_{lc}$.
\end{definition}

This work indirectly adjusts the distribution of latent confounders $Z_{lc}$ by leveraging the proxy variable $W$, enabling effective recovery of the latent structure even with low-quality proxies, thereby improving adaptability to real-world scenarios with noisy data.

\section{The Proposed LCDR Method}

This section provides a detailed description of our proposed LCDR method. First, we present the overview of the LCDR framework. Next, we explain how the iVAE (recovering latent representations requires proxy variables) is used to generate latent causal representations and how these representations are employed to constrain the latent representations generated by the VAE (recovering latent representations does not require proxy variables). Finally, we outline the framework of LCDR, including its objective function and the pseudocode of LCDR method.

\begin{figure*}
    \vspace{-40pt}
    \centering{}
    \includegraphics[width=6.5in]{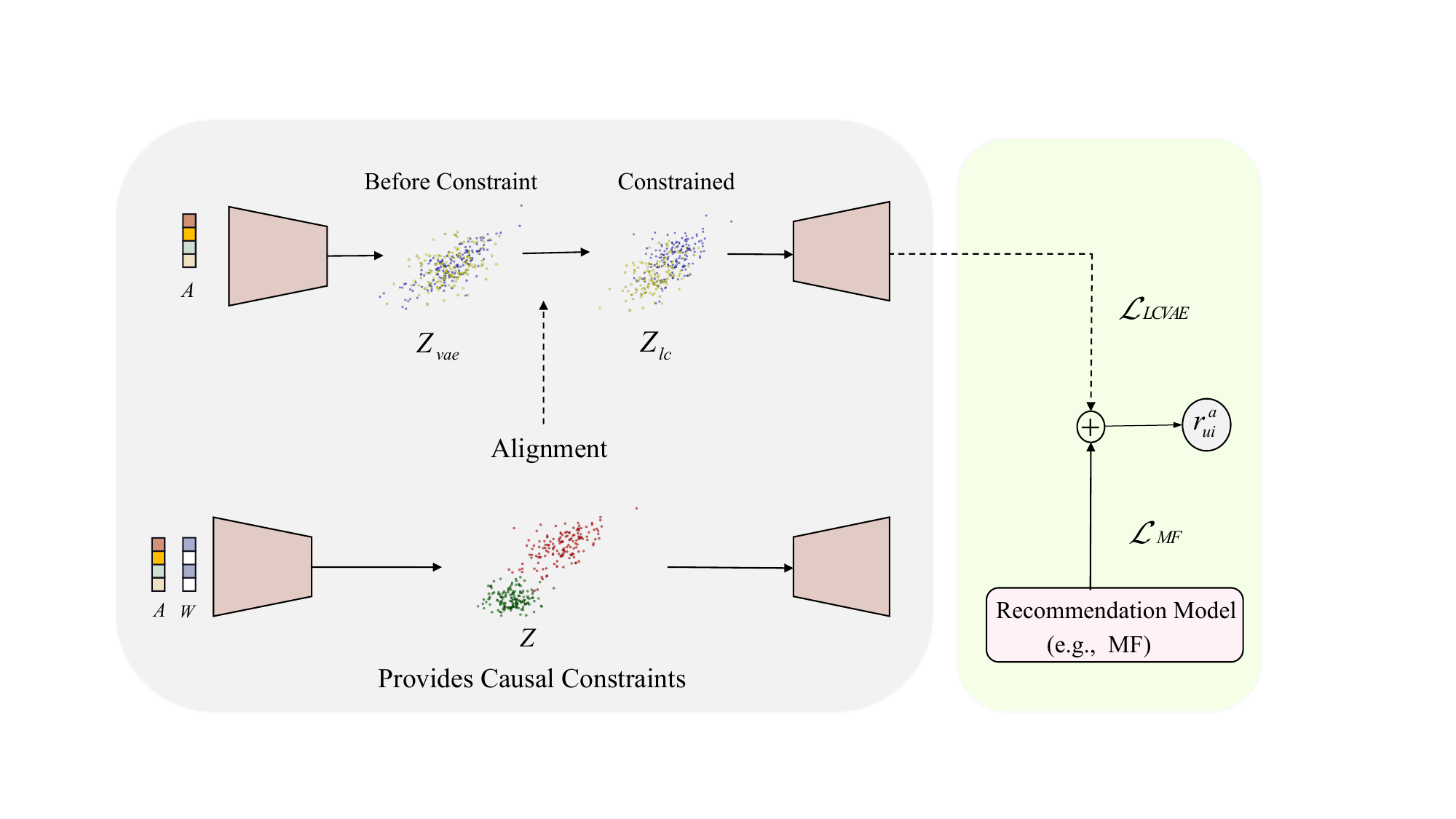} \vspace{-25pt}
    \caption{The overview of our proposed LCDR method. First, LCDR employs the latent causal representations $Z$, inferred by the iVAE, to constrain the representations generated by the VAE, resulting in the constrained representations $Z_{lc}$. Next, it leverages $Z_{lc}$ to enhance the performance of the recommendation model. This design is motivated by the fact that VAE cannot recover the true latent causal structure without proxy variables, and that proxy variables are often of low quality in practice.}
    \label{fig:1}
\vspace{-10pt}
\end{figure*}

\subsection{Overview of the LCDR Method}

The two main steps of the LCDR method are illustrated in Figure \ref{fig:1}.
In the first step, the representations generated by the VAE are aligned with the latent causal representations $Z$ learned from iVAE, thereby obtaining the constrained representations $Z_{lc}$. The second step focuses on enhancing the recommendation model by utilizing $Z_{lc}$.

\subsection{Generating Identifiable Latent Representations by iVAE}

We suppose that ${W}$ is a Bernoulli random variable with a mean $\mu({Z}) \in (0,1)$, where ${W}$ is correlated with ${Z}$ given the exposure status ${A}$.

Following the standard iVAE \cite{9},  we use $q_\phi\left({Z} \mid {A}, {W}\right)$  as the approximate posterior to learn $p_\theta\left({Z} \mid {A}, {W}\right)$:
\begin{equation}
\label{eq:13}
\small
\begin{split}
& E\left[\log p_\theta\left({A} \mid {W}\right)\right] \geq \mathcal{L}(\theta, \phi) \\
= & E[\underbrace{E_{q_\phi\left({Z} \mid {A}, {W}\right)}\left[\log p_\theta\left({Z} \mid {W}\right)-\log q_\phi\left({Z} \mid {A}, {W}\right)\right]}_I \\
& +\underbrace{E_{q_\phi\left({Z} \mid {A}, {W}\right)}\left[\log p_\theta\left({A} \mid {Z}\right)\right]}_{I I}]
\end{split}
\end{equation}

We can further decompose $\log p_\theta\left({A}, {Z} \mid {W}\right)$ into the following expression:
\begin{equation}
\begin{split}
&\log p_\theta\left({A}, {Z} \mid {W}\right)  \\=&\log p_\theta\left({A} \mid {Z}, {W}\right)+\log p_\theta\left({Z} \mid {W}\right) \\=&\log p_\theta\left({A} \mid {Z}\right)+\log p_\theta\left({Z} \mid {W}\right) 
\end{split}
\end{equation}

We use the Gaussian distribution as the prior distribution of the representations, defined as:
\begin{equation}
\begin{split}
p_\theta\left({Z} \mid {W}\right) & :=N\left(\mu_w\left({W}\right), v_w\left({W}\right)\right), \\
q_\phi\left({Z} \mid {A}, {W}\right) & := N\left(\mu_{a w}\left({A}, {W}\right), v_{a w}\left({A}, {W}\right)\right)
\end{split}
\end{equation}
where $N(\cdot, \cdot)$ is the Gaussian distribution. 
Thus, the computation of the expectation \(I\) from equation Eq.~\eqref{eq:13} can be simplified to the following equation:
\begin{equation}
\begin{aligned}
& E_{q_\phi\left({Z} \mid {A}, {W}\right)}\left[\log p_\theta\left({Z} \mid {W}\right) - \log q_\phi\left({Z} \mid {A}, {W}\right)\right] \\
&= -KL\left(N\left(\mu_{aw}\left({A}, {W}\right), v_{aw}\left({A}, {W}\right)\right), \right. \\
& \left. N\left(\mu_w\left({W}\right), v_w\left({W}\right)\right)\right)
\end{aligned}
\end{equation}

To compute the part $II$ in Eq.\eqref{eq:13}, we have the following equation:
\begin{equation}
\begin{split}
&\log p_\theta\left({A} \mid {Z}\right)=\sum_{i=1}^n a_{u i} \log \left(\mu_z\left({Z}\right)_i\right)+  \\& \left(1-a_{u i}\right) \log \left(1-\mu_z\left({Z}\right)_i\right)
\end{split}
\end{equation}

Thus, by optimizing Eq.\eqref{eq:13}, we can derive the approximate posterior $q_\phi\left({Z} \mid {A}, {W}\right)$ for recovering ${Z}$.

\subsection{Latent Causality Constraint VAE}
Following the standard VAE~\cite{vae}, we aim to learn the posterior distribution \(p_\theta(Z_{lc} \mid A)\) by introducing an approximate posterior distribution \(q_\phi(Z_{lc} \mid A)\), optimised using variational inference.

The objective function is defined as the Evidence Lower Bound (ELBO):
\begin{equation}
\begin{aligned}
\label{eq:vae1}
&\mathcal{L}(\theta, \phi) = \underbrace{{E}_{q_\phi(Z_{lc} \mid A)} \left[ \log p_\theta(A \mid Z_{lc}) \right]}_{I} - \\
& \underbrace{{KL}\left(q_\phi(Z_{lc} \mid A) \parallel p_\theta(Z_{lc})\right)}_{II} 
-\underbrace{\lambda*\left\|Z_{lc}-Z\right\|_2}_{III},
\end{aligned}
\end{equation}
where \(\log p_\theta(A \mid Z_{lc})\) is the log-likelihood of the generative model, \({KL}(\cdot\|\cdot)\) represents the Kullback-Leibler divergence measuring the discrepancy between the approximate posterior and the prior, and \(\| \cdot \|_2\) represents the $L_2$-norm (also known as the Euclidean norm).

The generative process can be decomposed as:
\begin{equation}
\begin{aligned}
&\log p_\theta(A, Z_{lc}) \\
&= \log p_\theta(A \mid Z_{lc}) + \log p_\theta(Z_{lc}),
\end{aligned}
\end{equation}
where \(p_\theta(Z_{lc})\) represents the prior distribution of the latent variable, typically set as a standard normal distribution \({N}(0, I)\). The term \(p_\theta(A \mid Z_{lc})\) denotes the conditional distribution of the observational data given the latent variable.

Furthermore, we make the following assumption:
\begin{equation}
\begin{aligned}
p_\theta(Z_{lc}) & := {N}(0, I), \\
q_\phi(Z_{lc} \mid A) & := {N}(\mu_\phi(A), \sigma_\phi^2(A)),
\end{aligned}
\end{equation}
where \(\mu_\phi(A)\) and \(\sigma_\phi^2(A)\) are parameters predicted by the encoder network.

To compute the term $I$ in Eq.\eqref{eq:vae1}, we have:
\begin{equation}
\begin{aligned}
&\log p_\theta(A \mid Z_{lc}) = \sum_{i=1}^n \biggl[a_{ui} \log(\mu_z(Z_{lc})_i) \\
&\quad + (1 - a_{ui}) \log(1 - \mu_z(Z_{lc})_i)\biggr],
\end{aligned}
\end{equation}
where \(\mu_z(Z_{lc})_i\) is the output of the decoder network.

The computation of term $II$ in Eq.\eqref{eq:vae1} is as follows:
\begin{equation}
\begin{aligned}
&{KL}\left(q_\phi(Z_{lc} \mid A) \parallel p_\theta(Z_{lc})\right) \\
&= \int q_\phi\left(Z_{lc} \mid A\right) \log \frac{q_\phi\left(Z_{lc} \mid A\right)}{p_\theta\left(Z_{lc}\right)} d Z_{lc}
\end{aligned}
\end{equation}

To compute the term $III$ in Eq.\eqref{eq:vae1}, we have:
\begin{equation}
\begin{aligned}
\left\|Z_{lc}-Z\right\|_2=\sqrt{\sum_{i=1}^n\left(Z_{{lc}, i}-Z_{i}\right)^2}
\end{aligned}
\end{equation}

By maximizing the Eq.\eqref{eq:vae1}, we can derive an approximate posterior 
\(q_\phi(Z_{lc} \mid A)\) for recovering ${Z_{lc}}$.
Then, the formula for calculating $\mathcal{L}_{LCVAE}$ is as follows:.
\begin{equation}
\mathcal{L}_{LCVAE} = E_{q_\phi\left({Z}_{lc} \mid {A}\right)}
\end{equation}

\subsection{Mitigate Confounding Bias between Item
Exposure and User Feedback}

Suppose that the conditional distribution $p({Z_{lc}} \mid {A}, {W})$ is uniquely characterised by the factor model~\cite{10.5555/120565.120567}, we now outline a general approach for identifying $p(r_{ui}^a)$ using proxy variables, as detailed below~\cite{20}:

\begin{equation}
\label{eq:1}
\begin{split}
&p(r_{u i}^a) 
= E_{{Z_{lc}}}[p(r_{u i} | {A}, {Z_{lc}})] 
\\&= \int_c p({Z_{lc}} = z) p(r_{u i} | {A}, {Z_{lc}} = z) dZ_{lc}.
\end{split}
\end{equation}

From Eq. \ref{eq:1}, it is clear that the key to estimating \( p(r_{ui}^a) \) lies in determining both \( p({Z_{lc}}) \) and \( p(r_{ui} \mid {A}, {Z_{lc}}) \). 
This process can be broken down into two steps: first, estimating \( p({Z_{lc}}) \), and second, determining \( p(r_{ui} \mid {A}, {Z_{lc}}) \).

\textbf{Step I:}
The first step is to obtain \( p({Z_{lc}}) \). 
This involves learning \( {Z_{lc}} \) from  \( {W} \), ensuring that the learned \( {Z_{lc}} \) closely approximates the true latent confounders \( {Z_{lc}} \) \cite{9,17} as follows:

\begin{equation}
\begin{split}
p({Z_{lc}} = z) &= E_{{A}, {W}}[p({Z_{lc}} | {A}, {W})]
\end{split}
\end{equation}

Hence, we need to learn $p\left({Z_{lc}} \mid {A}, {W}\right)$ from the data. In practice, we use the following equation:

\begin{equation}
\label{eq:3}
\begin{split}
&p({A} | {W}) = \int_c p({A} | {Z_{lc}} = z) p({Z_{lc}} = z | {A}, {W}) d Z_{lc}
\end{split}
\end{equation}
where $p({A} | {W})$ can be learned directly from the data. Therefore, we need to recover $p\left({Z_{lc}} \mid {A}, {W}\right)$ and $p\left({A} \mid {Z_{lc}}\right)$ from the data. 

\textbf{Step II:}
In the second step, we obtain $p\left(r_{u i} \mid {A}, {Z_{lc}}\right)$, which can be inferred from $p\left({Z_{lc}} \mid {A}, {W}\right)$ (learned in the \textbf{Step I} using iVAE) and $p(r_{u i} | {A}, W)$ as follows: 
\begin{equation}
\begin{split}
&p(r_{u i} | {A}, {W}) = \int_z p(r_{u i} | {A}, {Z_{lc}} = z) p({Z_{lc}} = z | {A}, {W}) d Z_{lc}
\end{split}
\end{equation}

With these two steps, we can obtain $p\left({Z_{lc}}\right)$ and $p\left(r_{u i} \mid {A}, {Z_{lc}}\right)$.
Eq.~\ref{eq:1} enables the identification of the potential outcome distribution $p\left(r_{u i}^a\right)$ based on observable data.

\subsection{Latent Causal-Constrained Debiasing Recommendation Framework}
For comparative purposes, we adopt the Matrix Factorisation (MF) model as the recommendation model in our framework, given its simplicity and widespread use in tasks involving the removal of latent confounding bias~\cite{23,29,idcf}.

The predicted rating, $\mathcal{L}_{MF}$, is formulated as follows:
\begin{equation}
\begin{split}
\mathcal{L}_{MF} ={P}_u^T {Q}_i + {B}_u + {B}_i
\end{split}
\end{equation} 
where ${P}_u$ denotes the latent vector for user $u$, ${Q}_i$ denotes the latent vector for item $i$, and $T$ represents the transpose operation. Additionally, ${B}_u$ represents the user preference bias term, while ${B}_i$ is the item preference bias term.

Finally, we utilise a point-wise recommendation model parameterised by \(\eta\) to estimate 
\(p\left(r_{ui} \mid {A}, {Z_{lc}}\right)\). Specifically, we adopt a simple additive model \(f(u, i, {Z_{lc}} ; \eta)\), which is defined as follows:
\begin{equation}
\begin{split}
\label{eq:18}
&f(u, i, {Z_{lc}} ; \eta) = \mathcal{L}_{LCVAE}  + \mathcal{L}_{MF}
\end{split}
\end{equation}

\subsection{The LCDR Algorithm}

The training process of LCDR is outlined in Algorithm~\ref{alg}, comprising two major steps:

The first step is learning latent causal-constrained representation $Z_{lc}$.
In this phase, the iVAE is used to infer latent causal representations, which serve as a constraint on the representations learned by a standard VAE. This alignment is achieved through a shared loss function, enabling the standard VAE to incorporate causal information even from noisy or weak proxy variables.
The second step is training the recommendation model.
Once the latent causal-constrained representations $Z_{lc}$ are obtained, they are fed into the recommendation model. The model parameters are then optimised using a task-specific loss (e.g., rating prediction or ranking loss), as defined in Eq.~\ref{eq:18}. 
 
\textbf{Time Complexity Analysis.}
The time complexity of LCDR consists of two phases: (1) learning the latent causal-constrained representation \( {Z}_{lc} \), and (2) training the recommendation model. In the first stage, both iVAE and LCVAE are executed in parallel to infer representations \( Z \) and \( {Z}_{lc} \), respectively. Let \( {C}_{iVAE} \) and \( {C}_{LCVAE} \) denote the per-sample computational cost of iVAE and LCVAE forward passes. Given a batch size and total training steps \( {T}_{1} \), the time complexity of this stage is \( \mathcal{O}\left( {{T}_{1} \cdot  B \cdot  \max \left( {{C}_{iVAE}, {C}_{LCVAE}}\right) }\right) \), since parallel execution is dominated by the slower branch. In the second stage, we train the recommendation model using \( {Z}_{lc} \). Let \( {C}_{f} \) denote the per-sample cost of the recommendation model, with batch size \( {B}^{\prime } \) and \( {T}_{2} \) training steps. The complexity of this stage is \( \mathcal{O}\left( {{T}_{2} \cdot  {B}^{\prime } \cdot  {C}_{f}}\right) \). Overall, the total time complexity of LCDR is:

\begin{equation}
\begin{split}
\mathcal{O}\left( {{T}_{1} \cdot  B \cdot  \max \left( {{C}_{iVAE}, {C}_{LCVAE}}\right)  + {T}_{2} \cdot  {B}^{\prime } \cdot  {C}_{f}}\right) 
\end{split}
\end{equation}

Although LCDR incorporates additional causal constraints (iVAE) into the representation learning process of VAE, its parallel architecture ensures that the computational overhead remains manageable. Consequently, LCDR effectively mitigates bias and enhances recommendation accuracy without incurring substantial training costs.

\begin{algorithm}
    \SetKwInOut{Input}{Input}\SetKwInOut{Output}{Output}
    \caption{LCDR Training}
    \label{alg}
    \Input{$\{{A}, {W}\}$, $\{r_{ui}\}, \forall (u,i) \in \mathcal{D}$} 
    \tcp{{Learning $Z_{lc}$}}
    \Repeat{convergence}{
    Sample a batch of interactions \(D \subset \mathcal{D}\)\;
    \For{each \((A, W) \in D\)}{
        \(\text{mean\_i}, \text{log\_var\_i} \leftarrow \text{iVAE}(A, W)\)\;
        
        \(Z \leftarrow \text{reparam}(\text{mean\_i},\text{log\_var\_i})\)\;
        
        \(\text{mean}, \text{log\_var} \leftarrow \text{LCVAE}(A)\)\;
        
        \(Z_{lc} \leftarrow \text{reparam}(\text{mean},\text{log\_var})\)\;

        

        Calculate loss as \( \text{loss} = -\mathcal{L}(\theta, \phi) \) by Eq.~\eqref{eq:vae1}\;

        Update \(\theta\) and \(\phi\) via gradient descent step on \(\nabla_{\theta, \phi} \text{loss}\);

    }
    }
    
   \tcp{Training recommendation model}
   
    Initialise a recommendation model $f(u, i,Z_{lc}; \eta)$ with parameters $\eta$\;
    \While{Stop condition is not reached}
    {
    Fetch $(u,i)$ from $\mathcal{D}$\;
    Minimise the loss Eq.\eqref{eq:18} to optimise $\eta$;
    }

    \Output{$\{\eta\}$}
\end{algorithm}

\subsection{The Identifiability of the Learned Representation}
In this section, we prove that the latent causal representation ${Z}$ learned by the iVAE component of LCDR is identifiable~\cite{9,xu2024causal}.

Let $\theta=(f, H, \lambda)$ be the parameter of the model $\Theta$, which defines the following conditional distribution:
\begin{equation}
\begin{split}
\label{zm1}
p_\theta\left({A}, {Z} \mid {W}\right)=p_f\left({A} \mid {Z}\right) p_{H, \lambda}({Z} \mid {W})
\end{split}
\end{equation}
where \( p_f\left({A} \mid {Z}\right) \) is defined as:

\begin{equation}
\begin{split}
\label{zm2}
p_f\left({A} \mid {Z}\right)=p_{\varepsilon}\left({A}-\mathbf{f}({Z})\right)
\end{split}
\end{equation}
where the value of \( {A} \) is decomposed as \( {A} = f({Z}) + \varepsilon \). Here, \( \varepsilon \) is an independent noise variable (independent of both \( {Z} \) and \( f \)), and its probability density function is given by \( p_{\varepsilon}(\varepsilon) \).


We impose the following assumption on the conditional distribution $p_{H, \lambda}({Z} \mid {W})$, modeling the relationship between the latent representation $Z$ and the proxy variables $W$:

\begin{assumption}
\label{assumption 1}
The conditioning on ${W}$ utilises an arbitrary function, like a look-up table or a neural network, that produces the specific exponential family parameters $\lambda_{i, j}$. Consequently, the probability density function is expressed as follows:
\begin{equation}
\begin{split}
\label{zm3}
& p_{H, \lambda}({Z} \mid {W})= \prod_i \frac{Q_i\left({Z}_{i}\right)}{Z_i({W})} \exp \left[\sum_{j=1}^k H_{i, j}\left({Z}_{i}\right) \lambda_{i, j}({W})\right]
\end{split}
\end{equation}
where $Q_i$ is the base measure, $Z_i({W})$ is the normalisation constant, and $\lambda_i({W})=\left(\lambda_{i, 1}({W}), \ldots, \lambda_{i, k}({W})\right)$ is a parameter related to ${W}$, $\mathbf{H}_i=\left(H_{i, 1}, \ldots, H_{i, k}\right)$ is the sufficient statistic, and $k$ is the dimension of each sufficient statistic, held constant. 
\end{assumption}

To establish the identifiability of the learned representation ${Z}$ in our LCDR, we introduce the following definitions \cite{9}:

\begin{definition} [Identifiability classes]
\label{Definition 2}
Let $\sim$ be an equivalence relation on $\Theta$. We say that Eq. \eqref{zm1} is identifiable (or $\sim$ identifiable) under $\sim$ if:
\begin{equation}
\begin{split}
p_{\mathbf{\theta}}({A})=p_{\tilde{\theta}}({A}) \Longrightarrow \tilde{\mathbf{\theta}} \sim \mathbf{\theta}
\end{split}
\end{equation}

The quotient space $\Theta / \sim$ is referred to as identifiability classes.
\end{definition}

We define an equivalence relation $\sim_M$ on the parameter space $\Theta$ as follows:
\begin{definition}
Let $\sim_M$ be the equivalence relation defined on $\Theta$ as follows:

\begin{equation}
\begin{split}
& (\mathbf{f}, \mathbf{H}, \mathbf{\lambda}) \sim(\tilde{\mathbf{f}}, \tilde{\mathbf{H}}, \tilde{\mathbf{\lambda}}) 
 \Leftrightarrow \exists \mathbf{M}, {Q} \mid \mathbf{H}\left(\mathbf{f}^{-1}({A})\right)
 \\&=\mathbf{M} \tilde{\mathbf{H}}\left(\tilde{\mathbf{f}}^{-1}({A})\right)+{Q}, \forall {A} \in \mathcal{X}
\end{split}
\end{equation}
where $\mathbf{M}$ is an $n k \times n k$ matrix and ${Q}$ is a vector.

\end{definition}

We now present our main result on the identifiability of the learned representations:

\begin{theorem}
\label{Theorem 2}
Suppose we have data collected from a generative model as defined by Eqs. \eqref{zm1}-\eqref{zm3}, with parameters \( (\mathbf{f}, \mathbf{H}, \mathbf{\lambda}) \). 
If the following conditions hold:
\begin{enumerate}[label=(\roman *)]
\item 
\label{i}
The set $\left\{{A} \in \mathcal{X} \mid \varphi_{\varepsilon}({A})=0\right\}$ has measure zero, where $\varphi_{\varepsilon}$ is the  characteristic function of the density $p_{\varepsilon}$ defined in Eq. \eqref{zm2}. 
\item 
\label{ii}
The hybrid function $\mathbf{f}$ in Eq. \eqref{zm2} is injective.
\item
\label{iii}
The sufficient statistic $H_{i, j}$ in Eq. \eqref{zm3} is differentiable almost everywhere, and $\left(H_{i, j}\right)_{1 \leq j \leq k}$ is linearly uncorrelated on any subset of $\mathcal{X}$ that has a measure greater than zero.
\item
\label{iv}
There exist $n k+1$ distinct points ${W}_0, \ldots, {W}_{n k}$ such that the matrix
\begin{equation}
\begin{split}
L=\left(\mathbf{\lambda}\left({W}_1\right)-\mathbf{\lambda}\left({W}_0\right), \ldots, \mathbf{\lambda}\left({W}_{n k}\right)-\mathbf{\lambda}\left({W}_0\right)\right)
\end{split}
\end{equation}
of size $n k \times n k$ is invertible.
\end{enumerate}
Then the parameters $(\mathbf{f}, \mathbf{H}, \mathbf{\lambda})$ are $\sim_M$ identifiable.
\end{theorem}

\paragraph{Limitations.}
Similar to other methods  based on iVAE (such as the previous SOTA method, iDCF), LCDR can leverage auxiliary observational variables to enhance identifiability when available, but these variables are not strictly required. Such auxiliary data (e.g., timestamps, user behavior sequences, demographic attributes) are commonly present in recommendation datasets. Importantly, LCDR retains functionality even without these variables—a key advantage over methods like iDCF that strictly depend on proxies. The inclusion of auxiliary variables (when measurable) further improves latent confounder recovery, but LCDR’s design ensures robustness to their absence.

\section{Experiments}

In this section, we conduct a comprehensive experimental study to address the following research questions:

\textbf{RQ1:}  How does the performance of our proposed model with latent causality constraints compare to other models that rely on ideal assumptions?

\textbf{RQ2:} Does our LCDR framework outperform a simple combination of the latent representations from VAE and iVAE?

\textbf{RQ3:} How do hyper-parameters influence the performance of LCDR?

\vspace{10pt}

\subsection{Experimental Settings}
We begin by providing an overview of the datasets, followed by a description of the baseline approaches and evaluation metrics. Finally, we outline the parameter configurations used for the LCDR model.

\paragraph{Datasets.}
To validate the effectiveness of our model, we conducted experiments on three publicly benchmark datasets. Table~\ref{t1} summarises the key characteristics of these datasets. The specific details of each dataset are as follows:

\textbf{Coat Dataset.} The Coat dataset \cite{coat} is a widely used benchmark dataset for evaluating  counterfactual recommendation algorithms. It contains data from 290 users selecting 300 coats. The dataset comprises  6,960 samples of coats chosen by the users themselves and 4,640 samples of coats chosen randomly by the users. Additionally, the dataset collects user feature information (such as gender, age, etc.) and coat feature information (such as type, color, etc.). User ratings are recorded on a 5-point scale, where a rating of 1 indicates the lowest level of preference (i.e., the user dislikes the coat the most).

\textbf{Yahoo!R3 Dataset.} The Yahoo!R3 dataset \cite{yahoo} is a dataset released by Yahoo containing user ratings for songs. It is widely used as a benchmark dataset in recommendation system research. The dataset collects 5-point scale ratings (where 1 represents the lowest score) from 5,400 users for 1,000 songs. It includes 129,179 MNAR (Missing Not At Random) samples (songs chosen by the users themselves) and 54,000 MCAR (Missing Completely At Random) samples (songs chosen randomly by the users). Additionally, the dataset provides features related to the likelihood of users rating the songs.

\textbf{KuaiRand Dataset.} The KuaiRand dataset \cite{gao2022kuairand} is a video dataset released by Kuaishou. It has been widely used in recommendation system research due to its extensive information collection. The dataset collects 1,413,574 MNAR samples (videos chosen by the users themselves) and 954,814 MCAR samples (videos chosen randomly by the users) from 23,533 users and 6,712 videos. The dataset also includes user features (such as registration time, number of followers, etc.) and video features (such as video type, playback duration, etc.). Furthermore, it provides signals indicating whether users clicked on the recommended videos.

Following iDCF~\cite{idcf}, we use all biased data for training, 30\% of the unbiased data for validation, and the remaining unbiased data for testing. 
"IsClick = 1" (KuaiRand) or ratings $\geq$ 4 (Coat and Yahoo!R3) as positive feedback.

\begin{table}
\centering
\caption{The statistic of Coat, Yahoo!R3, and KuaiRand.}
\label{t1}
\resizebox{0.48\textwidth}{!}{
\begin{tabular}{ccccc}
\hline Dataset & \#User & \#Item & \#Biased Data & \#Unbiased Data \\
\hline Coat & 290 & 300 & 6,960 & 4,640 \\
Yahoo! R3 & 5,400 & 1,000 & 129,179 & 54,000 \\
KuaiRand & 23,533 & 6,712 & $1,413,574$ & 954,814 \\
\hline
\end{tabular}
}
\end{table}

\begin{figure*}

    \centering
    \begin{minipage}[a]{0.45\linewidth}
        \centering
        \includegraphics[width=\linewidth]{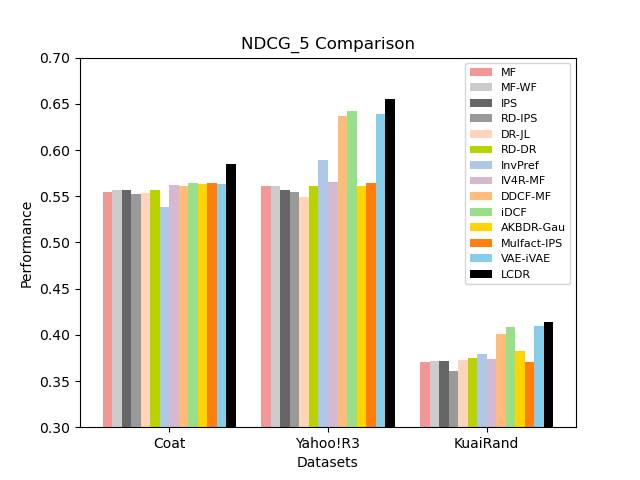}
         \captionsetup{labelformat=empty} 
    \end{minipage}
    \begin{minipage}[a]{0.45\linewidth}
        \centering
        \includegraphics[width=\linewidth]{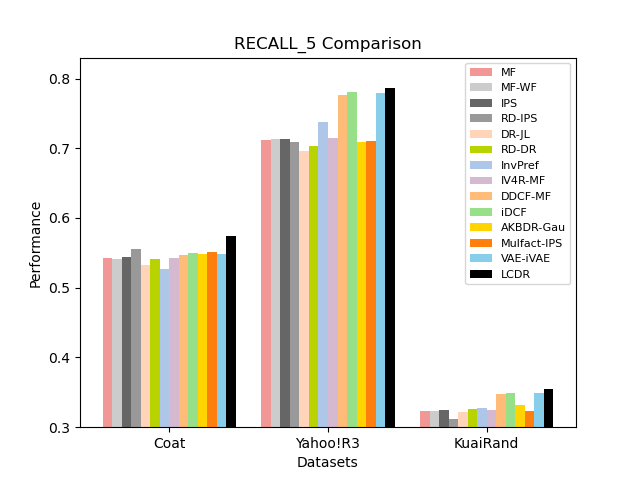}
         \captionsetup{labelformat=empty} 
    \end{minipage}
    
    \caption{The performance of all methods on the three real-world datasets.}
    \label{figp}   
    \vspace{-10pt}
\end{figure*}

\begin{table*}[!t]
\centering
\vspace{10pt}
\caption{The performance of all methods on the three real-world datasets is reported, with mean and standard deviation values calculated over ten runs. The best results are highlighted in boldface, while the runner-up results are underlined. Additionally, the p-value from a t-test comparing the performance of LCDR with the best-performing method for each dataset is provided.}
\resizebox{1\textwidth}{!}{
\label{tab:2}
\begin{tabular}{c|cc|cc|cc}
\hline
\multirow{2}{*}{Datasets} & \multicolumn{2}{|c|}{Coat} & \multicolumn{2}{c|}{Yahoo!R3} & \multicolumn{2}{c}{KuaiRand} \\
& NDCG@5 & RECALL@5 & NDCG@5 & RECALL@5 & NDCG@5 & RECALL@5 \\
\hline 
MF & $0.5546 \pm 0.0111$ & $0.5425 \pm 0.0116$ & $0.5610 \pm 0.0078$ & $0.7112 \pm 0.0072$ & $0.3705 \pm 0.0012$ & $0.3236 \pm 0.0013$ \\
MF-WF & $0.5566 \pm 0.0107$ & $0.5419 \pm 0.0112$ & $0.5608 \pm 0.0081 $ & $0.7136 \pm 0.0129$ & $0.3713 \pm 0.0013$ & $0.3239 \pm 0.0010$ \\
IPS & $0.5564 \pm 0.0124$ & $0.5442 \pm 0.0102$ & $0.5572 \pm 0.0038$ & $0.7131 \pm 0.0077$ & $0.3715 \pm 0.0013$ & $0.3242 \pm 0.0009$ \\
RD-IPS & $0.5549 \pm 0.0156$ & $0.5411 \pm 0.0153$ & $0.5543 \pm 0.0102$ & $0.7091 \pm 0.0123$ & $0.3609 \pm 0.0008$ & $0.3119 \pm 0.0013$ \\
DR-JL & $0.5531 \pm 0.0161$ & $0.5327 \pm 0.0108$ & $0.5487 \pm 0.0032$ & $0.6954 \pm 0.0046$ & $0.3727 \pm 0.0012$ & $0.3219 \pm 0.0010$ \\
RD-DR & $0.5568 \pm 0.0146$ & $0.5415 \pm 0.0117$ & $0.5609 \pm 0.0043$ & $0.7033 \pm 0.0061$ & $0.3749 \pm 0.0013$ & $0.3259 \pm 0.0009$ \\
InvPref & $0.5379 \pm 0.0117$ & $0.5272 \pm 0.0116$ & $0.5889 \pm 0.0038$ & $0.7382 \pm 0.0045$ & $0.3793 \pm 0.0014$ & $0.3276 \pm 0.0006$ \\
IV4R-MF & $0.5617 \pm 0.0099$ & $0.5433 \pm 0.0097$ & $0.5653 \pm 0.0062$ & $0.7152 \pm 0.0085$ & $0.3741 \pm 0.0010$ & $0.3251 \pm 0.0008$ \\
DDCF-MF & $0.5613 \pm 0.0096$ & $0.5467 \pm 0.0125$ & $0.6364 \pm 0.0061$ & $0.7762 \pm 0.0059$ & $0.4011 \pm 0.0009$ & $0.3469 \pm 0.0010$ \\
iDCF  & $0.5643 \pm 0.0114$ & $0.5498 \pm 0.0129$ & $\underline{0.6421} \pm 0.0027$ & $\underline{0.7806} \pm 0.0041$ & $0.4084 \pm 0.0005$ & $0.3485 \pm 0.0007$ \\
AKBDR-Gau & $0.5637 \pm 0.0106$ & $0.5489 \pm 0.0143$ & $0.5613 \pm 0.0039$ & $0.7087 \pm 0.0051$ & $0.3823 \pm 0.0011$ & $0.3314 \pm 0.0010$ \\
Mulfact-IPS  & $\underline{0.5646} \pm 0.0133$ & $\underline{0.5511} \pm 0.0135$ & $0.5648 \pm 0.0038$ & $0.7101 \pm 0.0073$ & $0.3702 \pm 0.0011$ & $0.3237 \pm 0.0012$ \\
VAE-iVAE  & $0.5632 \pm 0.0109$ & $0.5488 \pm 0.0116$ & $0.6389 \pm 0.0037$ & $0.7793 \pm 0.0033$ & $\underline{0.4092} \pm 0.0008$ & $\underline{0.3489} \pm 0.0007$ \\
LCDR  & $\mathbf{0.5973} \pm 0.0101$ & $\mathbf{0.5878} \pm 0.0119$ & $\mathbf{0.6631} \pm 0.0033$ & $\mathbf{0.7928} \pm 0.0034$ & $\mathbf{0.4176} \pm 0.0008$ & $\mathbf{0.3575} \pm 0.0011$ \\ 
\hline
p-value & $1 e^{-5}$ & $4 e^{-6}$ & $1 e^{-11}$ & $1 e^{-6}$ & $5 e^{-15}$ & $1 e^{-12}$ \\
\hline
 
\end{tabular}
}
\vspace{10pt}
\end{table*}


\paragraph{Baselines.}
To demonstrate the effectiveness of our LCDR method, we compare our approach with the following state-of-the-art (SOTA) deconfounding methods:
(1) \textbf{MF}~\cite{12} \& MF with features (\textbf{MF-WF}): MF is a classic collaborative filtering method that decomposes the high-dimensional rating matrix into the product of two low-dimensional matrices to capture latent features between users and items. MF-WF is an enhanced version of MF that incorporates user features to improve performance.
(2) \textbf{IPS}~\cite{23} \& \textbf{RD-IPS}~\cite{4}: IPS employs causal inference techniques to estimate propensity scores from observational data and uses IPW to mitigate selection bias. RD-IPS extends the IPS method by incorporating sensitivity analysis and adversarial learning to further optimise its performance.
(3) \textbf{DR-JL}~\cite{wang2019doubly}: DR-JL uses a dual robust estimator to mitigate the influence of missing not at random (MNAR) data in recommender systems.
(4) \textbf{RD-DR}~\cite{4}: RD-DR builds upon conventional DR methods by incorporating sensitivity analysis and adversarial learning to mitigate latent confounding bias.
(5) \textbf{InvPref} \cite{30}: InvPref uses invariant learning to separate users' true preferences from biased observational data, thereby reducing the influence of confounding bias.
(6) \textbf{IV4R-MF} \cite{IV-serach-data}: IV4Rec applies IVs to decompose input vectors within recommendation models to mitigate the effects of latent confounders. IV4R-MF uses MF as the backbone.
(7) \textbf{iDCF} \cite{idcf}: iDCF assumes that latent variables are independent and low-dimensional. It leverages proxy variables to infer identifiable latent confounders from observational data, which are subsequently used to mitigate confounding bias.
(8) \textbf{DDCF-MF} \cite{38}: DDCF assumes that no additional causal constraints are required when inferring latent confounders, and it employs deep learning techniques to infer latent confounders from observational data, which are then used to mitigate confounding bias. DDCF-MF is a variant of DDCF with MF as its backbone. 
(9) \textbf{AKBDR-Gau} \cite{li2024debiased}: AKBDR-Gau employs an adaptive kernel balancing method to find the optimal balancing function for mitigating selection bias.
(10) \textbf{Mulfact-IPS} \cite{huang2024going}: Mulfact-IPS assumes the absence of latent confounding factors and employs a propensity estimation method to address the popularity bias induced by item and rating factors.
(11) \textbf{VAE-iVAE}: For comparative purposes, we combine the latent representations learned by iVAE and VAE, using the resulting combined representations to enhance the recommendation model’s performance.

\paragraph{Evaluation metrics.} We evaluate the performance of each method using two widely adopted metrics in recommendation systems:  RECALL@K and NDCG@K, where K=5. Each method is tested across ten trials, and we report the average and standard deviation  to demonstrate the effectiveness and reliability of each method.

\paragraph{Implementation and Parameter Settings.}
We implement the LCDR framework using PyTorch and train it with the Adam optimiser~\cite{kingma2014adam}.
Grid search is employed to select the hyperparameters for all methods. The weight decay is chosen from  \{1e-5, 1e-6\}, while the learning rate is tuned from \{1e-3, 5e-4, 1e-4, 5e-5, 1e-5\}.
For the parameter $\lambda$ in Eq.\eqref{eq:vae1}, we set its value to 0.9 for Coat, 0.1 for Yahoo!R3, and 0.9 for KuaiRand.


\subsection{Performance Comparison (\textbf{RQ1})}



The experimental results are presented in Table \ref{tab:2} and Figure \ref{figp}. From the table, we observe the following:

\begin{itemize}
\item LCDR consistently outperforms other methods across all three datasets, demonstrating its ability to effectively impose constraints on latent causal representations during the generation of latent representations. This ensures that the model avoids overfitting to latent confounders and prevents the capture of non-causal noise. Furthermore, LCDR reduces reliance on assumptions such as the latent variable and the proxy variable are highly correlated, making it more aligned with the proxy variable are of low quality in real-world scenarios. 
\item Mulfact-IPS performs well on smaller datasets (e.g., Coat) but underperforms on larger datasets (e.g., KuaiRand).  This is likely due to its assumption of no latent confounders, which rarely holds in real-world scenarios. As the dataset size increases, the number of latent confounders also grows, leading to a significant decline in the performance of Mulfact-IPS compared to LCDR on larger datasets.
\item The performance limitations of IV4R-MF indicate that instrumental variables in real-world applications often fail to fully satisfy theoretical assumptions, thereby compromising model effectiveness.
\item iDCF underperforms compared to LCDR on all three datasets. This is because iDCF relies on idealised assumptions, such as the latent variable and the proxy variable are highly correlated, which are rarely satisfied in real-world scenarios (the proxy variable are of low quality in real-world scenarios). These assumptions limit its applicability. 
\item DDCF-MF also underperforms compared to LCDR on all three datasets. In real-world scenarios, without auxiliary proxy variables to assist in reconstructing the latent variable structure, models often struggle to accurately capture the true distribution of latent variables. Since DDCF-MF does not incorporate such auxiliary variables, its recovered latent variables are prone to noise interference and overfitting, ultimately limiting the model's performance.
\item (\textbf{RQ2}) VAE-iVAE performs worse than LCDR. Experimental results demonstrate that naively combining the latent representations of iVAE and VAE creates a critical issue: the proxy-assisted latent structure recovered by iVAE becomes contaminated by VAE's overfitted representations. In contrast, LCDR explicitly leverages the latent representations learned by iVAE to guide the representation learning in VAE. By using iVAE’s representations as a guiding signal, LCDR effective recovery of the latent structure even when the proxies are of low quality in real-world scenarios. 
\end{itemize}


\subsection{Ablation Study and Hyper-parameter Analysis} 

We conduct ablation studies and hyperparameter analysis to thoroughly investigate the impact of key components and parameter settings in the LCDR model on recommendation performance.

\paragraph{Ablation Study.}
To validate the effectiveness of the latent causal representation constraints in our model, we conducted ablation experiments by examining a variant of our LCDR model without latent causal representation constraints (LCDR w/o LC).

The results, presented in Table \ref{tab:4}, indicate that the model’s performance declines across all three datasets when the latent causal representation constraints are removed. This finding demonstrates that the latent representations generated by the VAE with latent causal constraints more accurately capture causal relationships in real-world scenarios. Furthermore, it highlights that latent causal constraints play a critical role in enhancing the performance of the recommendation model by mitigating overfitting to latent confounders and reducing the learning of non-causal noise.

\begin{table}[!t]
\centering
\caption{The ablation study of our LCDR method on three real-world datasets. The best results are in boldface, and the runner-up results are underlined.}
\label{tab:4}
\resizebox{0.48\textwidth}{!}{
\begin{tabular}{c|c|c|c}
\hline
Dataset & Method & NDCG@5 & RECALL@5 \\
\hline
\multirow{2}{*}{Coat} 
& LCDR w/o LC & $\underline{0.5612} \pm 0.0112$ & $\underline{0.5467} \pm 0.0133$ \\
& LCDR        & $\mathbf{0.5973} \pm 0.0101$ & $\mathbf{0.5878} \pm 0.0119$ \\
\hline
\multirow{2}{*}{Yahoo!R3} 
& LCDR w/o LC & $\underline{0.6409} \pm 0.0032$ & $\underline{0.7797} \pm 0.0039$ \\
& LCDR        & $\mathbf{0.6631} \pm 0.0033$ & $\mathbf{0.7928} \pm 0.0034$ \\
\hline
\multirow{2}{*}{KuaiRand} 
& LCDR w/o LC & $\underline{0.4069} \pm 0.0007$ & $\underline{0.3483} \pm 0.0007$ \\
& LCDR        & $\mathbf{0.4176} \pm 0.0008$ & $\mathbf{0.3575} \pm 0.0011$ \\
\hline
\end{tabular}
}
\end{table}

\begin{figure}
    \centering
    \begin{minipage}[a]{0.46\linewidth}
        \centering
        \includegraphics[width=\linewidth]{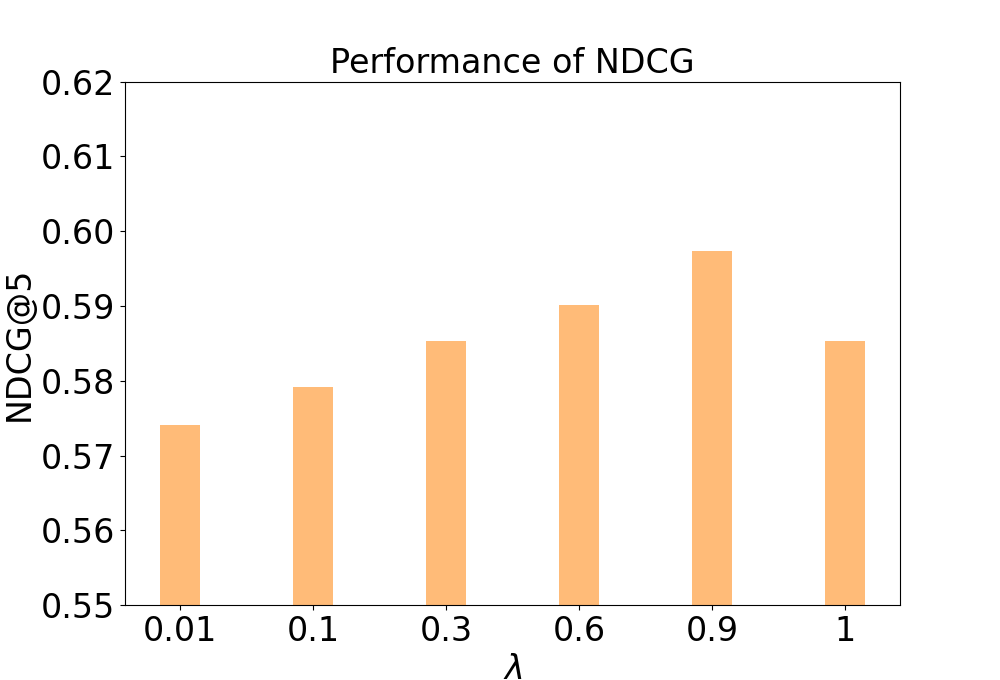}
         \captionsetup{labelformat=empty} 
        \caption*{(a) Effect of the $\lambda$ selection. We
        show the results of NDCG@5 on the Coat datasets.}
        \vspace{-0.3cm}  
    \end{minipage}
    \hfill
    \begin{minipage}[a]{0.46\linewidth}
        \centering
        \includegraphics[width=\linewidth]{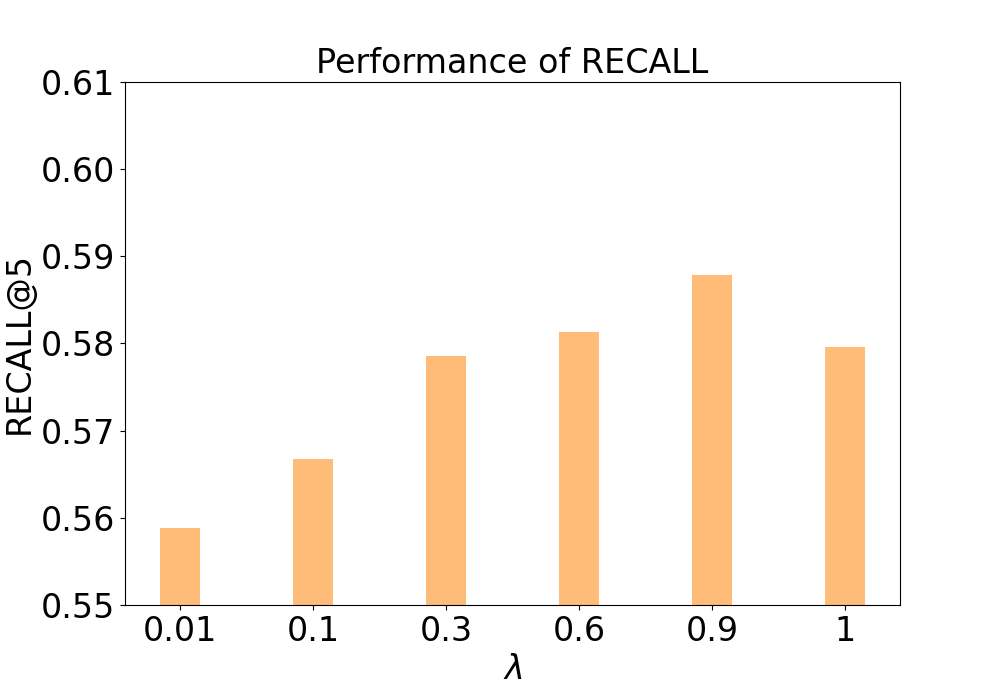}
         \captionsetup{labelformat=empty} 
        \caption*{(b) Effect of the $\lambda$ selection. We
        show the results of RECALL@5 on the Coat datasets.}
    \end{minipage}
    \hfill
    \caption{ Effect of the hyper-parameter selection.}
    \label{fighp}
\end{figure}
\paragraph{Hyper-parameter Analysis (\textbf{RQ3}).}
We conducted experiments to analyze the effect of the key hyper-parameter $\lambda$ in our LCDR method on recommendation performance.  We fixed all other parameters and varied $\lambda$ in Eq.\eqref{eq:vae1} from 0 to 1.
As shown in Figure \ref{fighp}, the results indicate the following:
As $\lambda$ increases, both NDCG@5 and RECALL@5 initially improve but then decline. This suggests that an appropriate level of latent causal constraints helps prevent the VAE from overfitting to latent confounders or learning non-causal noise, thereby improving performance. However, overly stringent causal constraints may cause the latent representations of the VAE to become overly similar to the low-dimensional representations learned by iVAE, resulting in a loss of complex causal information and a subsequent drop in performance.


\subsection{Computational Complexity Analysis} 

\begin{table}[!t]
\centering
\caption{A runtime comparison was conducted between LCDR and other two-stage latent confounding bias mitigation methods using the Coat dataset, yielding the following results.}
\label{tab:efficiency}
\begin{tabular}{l|c|c|c}
\hline
& \textbf{DDCF-MF} & \textbf{iDCF} & \textbf{LCDR} \\
\hline
Training (minutes) & 1.332 & 1.453 & 1.459 \\
Inference (ms/sample) & 0.407 & 0.411 & 0.415 \\
\hline
\end{tabular}
\end{table}

Table \ref{tab:efficiency} shows that compared with other two-stage latent confounding bias mitigation methods (DDCF-MF and iDCF), the training time of the LCDR model is almost the same as that of iDCF. This is because iVAE and VAE are trained simultaneously in LCDR.


\section{Conclusion}

In this work, we propose Latent Causality Constraints for Debiasing Representation Learning in Recommender Systems (LCDR) to address the limitations of existing debiased recommendation methods that rely on overly idealised assumptions, such as the availability of valid instrumental variables or strong proxy-confounder correlations. LCDR leverages the latent causal representations learned by iVAE to constrain the latent representations of VAE. This alignment enables the model to effectively utilize even weak or noisy proxy variables to recover latent confounders, thereby enhancing the performance of the recommendation model through the use of these constrained VAE representations. Extensive experiments on three real-world datasets demonstrate that LCDR consistently outperforms existing baselines in mitigating confounding bias and enhancing recommendation accuracy.

In the future, we plan to explore the potential of utilizing the LCDR model to address other challenges in recommender systems, such as selection bias.



\bibliographystyle{ACM-Reference-Format}
\balance
\bibliography{sample-base}

\end{document}